
\doublespace
 \newtoks\slashfraction
 \slashfraction={.13}
\def\lm{\lambda}
\def\kp{\kappa}
\def\shro{Schr\"odinger}
\def\apx{\approx}
\def\ll{\lbrack}
\def\rr{\rbrack}
\def\ch{{\cal H}}
\def\prt{\partial}
\def\pr{\prime}
\def\lmp{{\lm^\pr}}
\def\taup{{\tau^\pr}}
\def\hthing{{$H(\lm,\lm^\pr)$}}
\def\hthang{{H(\lm,\lm^\pr)}}
\def\dasl{{\it d}.{\it s}.{\it l}.}
 \def\slash#1{\setbox0\hbox{$ #1 $}
 \setbox0\hbox to \the\slashfraction\wd0{\hss \box0}/\box0 }
\pubnum={5377}
\pubtype={T}
 \date{February 1992}
\titlepage
\title{ $D=0$ Matrix Model as Conjugate Field Theory
 \foot{Work supported by the Department of Energy, contract
 DE-AC03-76SF00515.}}
\author{Shahar Ben-Menahem}
\SLAC
\medskip
\abstract
 The $D=0$ matrix model is reformulated as a nonlocal quantum
 field theory in two dimensions, in which the interactions occur
 on the one-dimensional line of hermitian matrix eigenvalues.
 The field can be thought of as a fluctuation
  in the potential $V$, and is conjugate to the density of
 matrix eigenvalues which appears in the Jevicki collective field
  theory. The classical solution of the field equation is either
 unique or labeled by a discrete index. Such a solution corresponds
to the Dyson sea modified by an entropy term. The modification
 smoothes the sea edges, and interpolates between different
  eigenvalue bands for multiple-well potentials. Our classical
   eigenvalue density contains nonplanar effects, and satisfies a
 local nonlinear Schr\"odinger equation with similarities to the
 Marinari-Parisi $D=1$ reformulation. The quantum fluctuations about
 a classical solution are computable, and the IR and UV divergences
 are manifestly removed to all orders. The quantum corrections
  greatly simplify in the double scaling limit, and include both
 string-perturbative and nonperturbative effects. The latter are
 unambiguous for $V$ bounded from below,
  and can be compared with the various nonperturbative definitions
 of these theories proposed in the literature.
\submit{Nuclear Physics \bf B}
\endpage
\chapter{Introduction.}
The study of discretized two-dimensional quantum gravity (matrix models)
has received much attention in recent years,
 thanks to the uncovering of the
solvability of a range of such models in the double scaling limit
\REFS\doug{M. Douglas and S. Shenker \journal Nucl.Phys.&B335(90)635
.}\REFSCON\grossmig{D.J. Gross and A.A. Migdal\journal
 Phys.Rev.Lett.&64(90)717.}\REFSCON
 \brezkaz{E. Brezin and V.A. Kazakov\journal
 Phys.Lett.&B236(90)144.}\REFSCON\brzkzzm{E. Brezin,
 V.A. Kazakov and A.B. Zamolodchikov,
 \journal Nucl.Phys.&B338(90)673.}
\REFSCON\grossmil{D.J. Gross and N. Miljkovic\journal
Phys.Lett.&B238(90)217
.}\REFSCON\pgins{P. Ginsparg and J. Zinn-Justin\journal Phys.Lett.
&B240(90)333.}\REFSCON\grossklb{D.J. Gross and I.R. Klebanov
 \journal Nucl.Phys.&B344(90)475
.}\REFSCON\klbwlk{I.R. Klebanov and R.B. Wilkinson,
Princeton preprint PUPT-1188(1990).}
\refsend.
 A matrix model describes the statistical mechanics of random surfaces
 in the large-$N$ limit, where $N$ is the dimension of the matrix
(we restrict our attention to hermitian matrix models).
 The procedure for solving these models consists of
three conceptual steps. In the first step, integrating out the
 angular modes of the random matrix leaves us with its
$N$ real eigenvalues as the dynamical variables. Then, the large-N limit
is performed via a WKB-like procedure\foot{For the $D=1$ model, this
is a true quantum-mechanical WKB procedure for $N$ independent fermions
in a potential well.}. And finally, as $N$ is taken to infinity, the
matrix-model potential must approach one of its critical values
in accordance with the double-scaling limit.
 The averages of various quantities over random surfaces
 can then be found
as functions of the string coupling, which is held fixed in this limit.
 To any given order in the string-perturbative
  expansion, the matrix model
results may be compared with
results from the corresponding continuum models,
which consist of the quantum Liouville theory coupled to various matter
fields.
\REFS\kpz{V. Knizhnik, A. Polyakov and A. Zamolodchikov
\journal Mod.Phys.Lett.&A3(88)819}
\REFSCON\ddk{F. David\journal Mod.Phys.Lett.&A3(88)1651;
\nextline J. Distler and H. Kawai
\journal Nucl.Phys.&B321(89)509.}\refsend
\REFS\plch{J. Polchinski, Texas preprint UTTG-19-90.}
\REFSCON\polch{J. Polchinski\journal Nucl.Phys.&B346(90)253
.}\REFSCON\yng{Z. Yang\journal Phys.Lett.&B255(91)215
.}\REFSCON\wise{A. Gupta, S.P. Trivedi
 and M.B. Wise
\journal Nucl.Phys.&B340(90)475.}\refsend.
 Such comparisons have been carried out for low genus
\REFS\kst{I. Kostov\journal Phys.Lett.&B215(88)499.}
 \REFSCON\newm{D.J. Gross, I.R. Klebanov and M.J. Newman\journal
  Nucl.Phys.&B350(91)621.}
\REFSCON\brshd{M. Bershadsky and I.R. Klebanov\journal
 Phys.Rev.Lett.&65(90)3088.}
 \REFSCON\SBM{S. Ben-Menahem\journal Nucl.Phys.&B364(91)681.}
 \refsend .
\par
 In addition to perturbative series in the string coupling,
 one finds in these models also
nonperturbative effects\REFS\npt{S. Shenker, Rutgers preprint
 RU-90-47(90), and references within.}\refsend.
 In the $D=1$ model, the leading such effect comes
from the tunneling of a single matrix eigenvalue (=fermion ) out of
the potential well, or between different potential wells.
The contribution to, say, the free energy will behave as
 $\exp(-{\it const
}/g_{string})$ for low string couplings.
 In ref.\npt\ it
 was pointed out that for the $D=0$ models, as well,
  there are saddle-point
 configurations where a single eigenvalue leaves the potential
well\foot{In the $D=0$ case, this is the effective potential, namely
the matrix-model potential plus the mean potential due to the Coulomb
repulsion of the other N-1 particles.}, and that these configurations
 give rise to nonperturbative effects of the same form. However, for
even-$k$ multicritical $D=0$ models, including pure gravity itself,
 there are ambiguities in defining the theory nonperturbatively, which
 can be traced to the fact that the critical quartic potential is
 unbounded from below. Various nonperturbative definitions
 have been proposed in the literature
\REFS\mira{J.Luis Miramontes and J.S. Guillen, CERN-TH. 6323/91.}
\REFSCON\dav{F. David\journal Mod.Phys.Lett.&A5(90)1019}
\REFSCON\fdav{F. David\journal Nucl.Phys.&B348(91)507}
\REFSCON\halp{J. Greensite and M.B. Halpern
\journal Nucl.Phys.&B348(91)507}
\REFSCON\marinar{E. Marinari and G. Parisi
\journal Phys.Lett.&B240(90)375}
\REFSCON\amb{J. Ambjorn, J. Greensite and S. Varsted
\journal Phys.Lett.&B249(90)411}
\REFSCON\marek{M. Karliner and A.A. Migdal\journal Mod.Phys.Lett.
 &A5(90)2565}
\REFSCON\ambb{J. Ambjorn, C.V. Johnson and T. Morris, Southampton
and Niels Bohr Inst. Preprint SHEP 90/91-29, NBI-HE-91-27.}
\REFSCON\dall{S. Dalley, C.V. Johnson and T. Morris, SHEP 90/91-28
and SHEP 90/91-35.}
\REFSCON\deo{K. Demeterfi, N. Deo, S. Jain and C.-I. Tan\journal
 Phys.Rev.&D42(90)4105.}
\REFSCON\bhanot{G. Bhanot, G. Mandal and O. Narayan\journal
 Phys.Lett.&B251(90)388.}
\REFSCON\josh{J. Feinberg, TECHNION-PH-92-1.}
\refsend
. For the even-$k$ multicritical models, these definitions can and
 do disagree with one another, since the perturbative series is
 not Borel-summable. But even for the well-defined models,
  a systematic derivation of the nonperturbative physics
  directly in $D=0$ has so far been lacking.
 \par
 A further line of development has been the reformulation of
the $D=1$ matrix model as a string field theory.
In matrix language, the correlators that one calculates in this model
 are of any
number of boundary operators, each such boundary having an arbitrary
length. Each boundary is in a single slice of the embedding dimension
\foot{Otherwise, the angular matrix variables are
 excited and the calculations
cannot be done using current methods.}.
  The same information is contained in the $n$-point functions
of the density of matrix eigenvalues; it is essentially
 this density which has been
suggested as the string field\REFS\dj{S.R. Das and A. Jevicki\journal
 Mod.Phys.Lett.&A5(90)1639.}\refsend .
 It is a function of
$\lambda$, the matrix eigenvalue, and the embedding dimension; hence
the field theory is two-dimensional. In ref.\ \lbrack\dj\rbrack,
 Das and Jevicki used
collective coordinate techniques to transform the quantum mechanics
of $N$ particles (in a bosonic formulation) into two-dimensional
quantum field theory. The kinetic term in the action is that of
a massless field, which corresponds to the massless tachyon
in the continuum effective field theory\ \lbrack\polch\rbrack.
 Some aspects of this
correspondence remain unclear--- for instance, the identification of
matrix eigenvalue with (a function of the) Liouville zero-mode.
Nevertheless, much progress has been made in understanding the
Das-Jevicki collective field theory, making string-perturbative
calculations with it, and comparing the results with those obtained
via other methods
\REFS\kres{K. Demeterfi, A. Jevicki and J.P. Rodrigues\journal
 Mod.Phys.Lett.&A6(91)3199
 , and references therein.}
\REFSCON\karab{D. Karabali and B. Sakita\journal Int.J.Mod.Phys.
&A6(91)5079.}\refsend
. Nonperturbative effects, appearing in the form of solitons and
instantons of the $D=1$ field theory, have also been investigated
\REFS\jevnpt{A. Jevicki, BROWN-HET-807.}
\REFSCON\avan{J. Avan and A. Jevicki\journal Phys.Lett.&B272(91)17.}
\refsend.
\par
Another approach to $D=1$ field theory has been to retain its
formulation as $N$-fermion quantum mechanics, but second-quantize
the fermions and then bosonize them\REFS\wad{A.M. Sengupta and S.R.
 Wadia\journal Int.J.Mod.Phys,&A6(91)1961
 .}\REFSCON\moor{
G. Moore, RU-91-12.}\REFSCON\sncn{D.J. Gross and I.R. Klebanov\journal
 Nucl.Phys.&B359(91)3.}
\refsend .
  This
approach is equivalent to the Das-Jevicki formulation.
\par
 Although the collective-field method has also been applied to the
 $D=0$ matrix models\REFS\jevsak{A. Jevicki and B. Sakita\journal
 Nucl.Phys.&B185(81)89.}\REFSCON\ajev{A. Jevicki\journal
Nucl.Phys.&B146(78)77.}\REFSCON\sakbook{B. Sakita, "Quantum Theory
 of Many-Variable Systems and Fields", World Scientific, Singapore,
1985.}\REFSCON\cohn{J.D. Cohn and S.P. de Alwis,
 COLO-HEP-247, IASSNS-HEP-91/7.}\refsend,
 it seems to us less compelling there than for $D=1$, because of its
 unusual kinetic term, and although a perturbative scheme has been
 outlined\REFS\jevsev{A. Jevicki, BROWN-HET-777.}
\REFSCON\reneg{O. Lechtenfeld, IASSNS-HEP-91/86.}
 \refsend , this
 program has not, to our knowledge, been carried out as far as the
 corresponding $D=1$ field theory\foot{In ref.\ \ll\reneg\rr\ ,
 Lechtenfeld uses our equations to set up a perturbative scheme. But
 since he performs the genus expansion at an early stage, one is left
 with manifest divergences, and in addition nonperturbative
 information is lost. We avoid these problems, as will be seen below.}.
  \par
  In this paper, we develop an alternative
   field theory formalism for the
 $D=0$ matrix models.
 Our field theory is two-dimensional, but the extra dimension
is an auxiliary one and drops out of the formalism at some stage.
 What is left is a nonlocal quantum mechanics, with a
  function of $\lambda$
playing the role of time. This is interesting in view
of the idea \lbrack\dj\rbrack\
  that the matrix eigenvalue is related to the Liouville zero-mode.
 The Jevicki-Sakita $D=0$ collective field theory is also a nonlocal
 quantum mechanics. We have found, however, that our formalism has the
 following desirable properties, which make it worth pursuing:\par
 {\bf A.\ } The entropy term\foot{This term is $N\int\rho\ln\rho
\; d\lambda$, with $\rho(\lambda)$ the eigenvalue density.} in the
action, which in the collective field approach appears through the
 Jacobian, is for the first time incorporated into the classical
 solution. This results in smoothing of the Dyson sea edges,
interpolation between different eigenvalue bands for multiple-well
 potentials (determining their relative population),
  and an unambiguous treatment
  of nonperturbative (instanton) effects, directly in a $D=0$
 field-theory framework, for multicritical models that can
 occur for a bounded-from-below potential.
 \par {\bf B.\ } \ Our classical eigenvalue
 distribution, $\rho(\lm)$, satisfies a local nonlinear Schr\"odinger
 equation, with $1/N$ playing the role of $\hbar$. The Schr\"odinger
 potential, $V_1(\lm)$, appearing in this equation is similar to,
 but distinct from,
 the Marinari-Parisi $D=1$ reformulation. On the other hand,
  in the planar limit it has an exact corresposondence with
  the Dyson sea effective potential, mentioned above.
\par {\bf C.\ } Not only are quantum corrections computable, but most
 of them vanish in the double scaling limit. The only
 quantum corrections which survive this limit, apart from the
 semiclassical functional determinant, are a set of Feynamn graphs
 which can be exactly summed.\medskip
 The rest of this paper is organized as follows. In section 2, the
matrix-model partition sum is recast as that of a massless
two-dimensional field, $A(r)$, with (nonlocal) interactions confined
 to an infinite line (the `eigenvalue axis'). The equation of motion
 is written, and its classical solution is expressed as an eigenvalue
 distribution, $\rho(\lm)$, and shown to be either unique or labeled
 by a discrete index. The two-dimensional equation of motion becomes
 a one-dimensional integro-differential equation on the eigenvalue
 axis; this in turn leads to a weaker\foot{`Weaker' because the
 Schr\"odinger equation has more solutions than the integro-differential
 equation.} local nonlinear Schr\"odinger equation. The interpretation
 of this equation is discussed, as well as the nature of its solution
 and how it is determined uniquely (up to the possible discrete index).
\par In section 3, the quantum fluctuations about a classical solution
 are studied, and an expression derived for the partition sum in terms
 of the propagator of the $A$ field (the `conjugate field'). The IR
 and UV singularities are seen to cancel in a trivial way, to all
 orders. The one-dimensional Green's equation for the propagator is
presented, as well as a partial solution. Section 4 is devoted to
 a discussion of the double scaling limit, including nonperturbative
 effects\foot{Most of the discussion in section 4 is limited to
 the case of the quartic potential as it approaches its $k=2$
  critical
  point. Since neither the theory nor our formalism are well defined
 for this potential, only the string-perturbative series discussed
 in that section is meaningful. Therefore, our description of
  an instanton
 ansatz for this model, described in sec.4,
 is not necessarily more reliable than any of
 the previous nonperturbative definitions of pure gravity
  referenced above.
 Nevertheless, it is reassuring that our exponentially-suppressed
tunneling factor, eq.(39a), agrees with that appearing in other
 approaches
  (it is the prefactor that is ambiguous). The main point
 of our instanton calculation is, however, that the conjugate-field
 formalism applies to any well-defined realization of a critical model,
 and a similar instanton calculation for such a realization would be
 rigorous.}.
  We see there that the quartic $k=2$
  classical solution is unique at
 the level of string perturbation theory. In section 5 we restate our
 conclusions.
  Some mathematical
 details are reserved for the appendix. Many results, presented or
 stated without proof in this paper, will be exposed more fully in
 a follow-up publication, to appear soon.\par
Finally, an explanatory note is in order. An early version of this
 preprint, containing most of the results (with the notable exception
 of point {\bf C} above), has been circulating privately since November
 1990. That earlier version has sometimes been confused in the
 literature with SLAC-PUB-5262, an altogether distinct work.\par
\chapter{Conjugate Field Formalism and Classical Solutions.}
 The partition function of the $D=0$ matrix model is,
$$Z\{ V\}\equiv\int\ll d\lm\rr \exp\bigl(-N\sum_{i=1}
^NV(\lm_i)+\sum_{i\not=
j}\ln\vert\lm_i-\lm_j\vert\bigr)\eqno (1)$$

Where $V$ is the matrix-model potential, which we assume to be
    a polynomial.

 We may formally rewrite ($C$ a divergent number)

$$\exp\bigl(\sum_{i\not= j}\ln\vert\lm_i-\lm_j\vert\bigr)=
C\int\ll dA\rr \exp\bigl(
 \sqrt{4\pi}i\sum_{j=1}^NA(\lm_j)-{{1}\over{2}}\int(\prt A)^2d\lm dx
 \bigr)
\eqno (2)$$
Which is the path integral over a massless field $A$ in two dimensions,
 with $N$ point charges on the $x=0$ (`eigenvalue') axis.
 $x$ is an auxiliary dimension, and $A(\lm)\equiv A(\lm 0)$.
We shall use $r$ to denote a general point $(\lm,x)$ in the plane.
 \par This path integral is beset by infrared and ultraviolet divergences.
 The $UV$ divergence is regulated by smearing the point charges;
$A(\lm_j)$ is replaced by
 $${1\over{\epsilon}}\int_{\lm_j}^{\lm_j+
\epsilon}d\lm A(\lm)\eqno(2a)$$
 in eq. (2). The infrared
  divergence is regulated by
 introducing a uniformly charged circle in the plane, centered about
 the origin and with a large radius $L$. The total charge of this
 circle is $-N$, which screens the $N$ point-charges.
  All dependences on $\epsilon$ and
$L$ will cancel in a simple way.
 In what follows, we will mostly ignore the need to regulate the path
integral; a more careful treatment reveals that this naive approach
is the correct one.
  Introducing the normalized charge density,

 $$\rho(\lm)={1\over{N}}\sum_i\delta(\lm-\lm_i)\eqno (3)$$
 We may instead think of $Z$ as a path integral over the overcomplete
 variables $\{\rho(\lm)\}$, with integrand

 $$\exp\bigl(-N^2\int d\lm\rho(\lm)V(\lm)+N^2\int\int
d\lm d\mu\rho(\lm)\rho(\mu)\ln\vert\lm-\mu\vert\bigr)\eqno (4)$$

This is the collective-field approach of Jevicki and Sakita. But we
 prefer to use the conjugate field $A(r)$ as our dynamical field,
 since it has a standard kinetic term and trivial Jacobian. The other
 merits of the conjugate field theory have been listed in points
 {\bf A, B} and {\bf C} of the introduction.\par
Combining eqs. (1) and (2), and using (2a) to regulate the UV
 divergence in $C$, we obtain the following expression for the
 partition function\foot{We define the measure $\ll dA\rr$ to include
 the infinite, but $\{L,\epsilon,N,V(\lm)\}$ independent, factor
 of $(\det^\prime\partial^2)^{1/2}$.}:
$$Z\{V\}=C_1(L,N)\epsilon^{-N}\int\ll dA\rr\bigl(I\{A\}\bigr)^N
\exp\bigl(-{i\over{\sqrt{\pi}}}{N\over L}\ointop dsA-{1\over 2}\int
(\prt A)^2d^2r\bigr)\eqno(5)$$

$$I\{A\}\equiv\int d\lm \exp\bigl(-NV(\lm)+i\sqrt{4\pi}A(\lm)\bigr)
  \eqno (6)$$
In equation (5), the linear term is an integral over the charged circle,
 and $C_1$ is IR divergent, UV finite, and $V$-independent. All integrals
 over $\lm$, here and below, range over the whole real axis, unless
 the limits of integration are explicitly indicated.\par The euclidean
 action is, apart from the IR-regulator source term,

  $$S\{A\}\equiv {1\over 2}\int(\prt A)^2d^2r-N\ln I\{A\}.\eqno(6a)$$

This is a nonlocal action, with the interactions occuring on the
eigenvalue axis. Away from the axis, $A$ is a free massless field.

$i\sqrt{4\pi}A/N$ can be thought of as a quantum fluctuation in
 the matrix-model potential.
 In addition, $A$ is conjugate to $\rho$ in the thermodynamical sense.
\par The classical
 equation of motion for the field $A$ is
\foot{ We omit a step here, related to the infrared
 regulator: the field $A$ should be shifted by a constant, which is the
constant potential due to the charged circle in its interior.}

  $$\prt^2A(r)=-i\sqrt{4\pi}N\rho_1(\lm)\delta(x)\eqno(7a)$$
where
  $$\rho_1(\lm)\equiv{1\over I}\exp\bigl(-NV(\lm)+
  i\sqrt{4\pi}A(\lm)\bigr) \eqno(7b)$$

 Note that although the action is nonlocal, the equation of motion
 is local, but with a new coupling $I$, to be determined
 self-consistently.\par
The boundary condition for the conjugate field $A$ at infinity is,
$$A(r)\approx -i{{2N}\over{\sqrt{4\pi}}}\ln\vert r\vert+\; const
\eqno (7c)$$
 since that is the electrostatic potential far from the $N$
 point-sources (eq.(2)). This boundary condition allowed
 us to freely integrate by parts the variation of the kinetic term
 in the action.\par
 We denote a solution of eqs.(7) by
$$A_{classical}(r)\equiv -iA_s(r)\eqno (7d)$$
 and define the classical charge (eigenvalue) density
  on the eigenvalue axis as $\rho$ at the classical solution:
  $$\rho(\lm)\equiv{1\over I_s}\exp\bigl(-NV(\lm)+
  \sqrt{4\pi}A_s(\lm)\bigr) \eqno(8)$$

Where $I_s$ is the value of $I\{A\}$ at the classical solution.

 Henceforth we shall adopt the definition eq.(8) for $\rho$, instead
 of eq.(3). From here on we shall work with $\rho$ instead of $A_s$;
 They contain the same information, although $\rho$ is defined only
 on the eigenvalue axis\foot{Just as, in the analog 2d electrostatic
 problem, the charge distribution on a plate together with the
boundary condition at infinity, determine the potential throughout
 an otherwise empty space.}
 . \par
Let us assume that the potential $V$ has been chosen to be bounded
 from below. In that case, as will be shown, $\rho(\lm)$ is unique
 or labeled by a discrete index. In section 4, we shall see that
 quantizing $A(r)$ about the classical configuration determined by
 $\rho$, gives rise to a unique string-perturbative expansion for
 physical quantities in the double scaling limit (abreviated henceforth
 as {\it d}.{\it s}.{\it l}.
 \foot{In order to properly analyze
  the d.s.l., a multiple-well
 potential, bounded from below, must be used. In this paper, when we
 specialize to a particular $V(\lm)$, it is the quartic
  $\lm^2/2+g\lm^4$, which is {\it not} bounded from below at
 criticality. This is not important for string perturbation theory;
  work on nonperturbative
 effects in which a multiple-well potential {\it is} used,
 is currently in progress.  }
 ), {\it provided} $\rho$ has support in a single Dyson sea (band)
 in the planar limit.
  Thus, the possible discrete non-uniqueness of $\rho$ will
 be revealed either
  at the nonperturbative level, or for multiband solutions.
 However, any {\it continuous} non-uniqueness of multiband solutions,
 which results from the freedom to adjust the relative populations of
  different bands, is eliminated by the nonperturbative effects
 (`tunneling'), as we shall see.
  \par
 Both the equation of motion and boundary condition for $A_s(r)$ are
 real, and thus each $\rho(\lm)$ in the discrete set of solutions
is either real, or the complex conjugate of another solution.
 The possible imaginary
 parts of solutions $\rho$ can only be revealed nonperturbatively;
 in any case, all physical quantities are real.    \par
 Evaluating the action at the classical solution, we find that the
  $L$-dependent prefactor in eq. (5) gets canceled:
 $$\left.\eqalign{
   Z_{\scriptstyle classical}\{V\}=&const\;\epsilon^{-N}\exp\bigl(
  -N^2\int d\lm\rho V\cr
 &+N^2\int\int d\lm d\mu\rho(\lm)\rho(\mu)\ln\vert\lm-\mu\vert-
       N\int d\lm\rho \ln\rho\bigr)}\right.\eqno(5a)$$

 The constant depends only on $N$, and is hence irrelevant.

 As we discuss below, the $UV$ divergence in eq (5a) is cancelled
 by the semiclassical (determinant) factor, whereas the higher
 quantum corrections to $Z$ are both $IR$- and $UV$-finite.

 \par
Notice the $O(N)$ correction to the free energy that occurs already
at the classical level. This correction has a simple interpretation:
It is a combinatorical factor, the entropy of the classical
 solution (see refs. \ll\jevsak\rr,\ll\ajev\rr,\ll\sakbook\rr,
 \ll\cohn\rr) .
Of course, the sum of all $O(N)$ corrections to the free energy must
vanish, since only even powers of $N$ appear in the topological genus
expansion\REFS\biz{D. Bessis, C. Itzykson and J.B. Zuber\journal
 Advances in Applied Math.&1(80)109.}\refsend.
\par

 $\rho$ satisfies the integral equation,

$$V(\lm)+{1\over N}\ln(\rho(\lm)I_s)=
2\int d\mu \ln\vert\lm-\mu\vert\rho(\mu)
             \eqno(9a)$$

 which follows from eqs.(7), in conjunction with the fact that the
 $2d$ free Green's function is ${1\over {2\pi}}\ln\vert r-r^\pr\vert$.
Upon differentiation w.r.t. $\lm$, eq.(9a) becomes the following
one-dimensional integro-differential equation:
$$V^\pr(\lm)+{1\over N}{{\rho^\pr(\lm)}\over{\rho(\lm)}}=2{\cal H}(\rho
(\lm))\eqno (9b)$$
Where ${\cal H}$ denotes the Hilbert transform:
$${\cal H}(f(\lm))\equiv\int{{f(\mu)d\mu}\over{\lm-\mu}}\eqno (9c)$$
 for any function f\foot{Such integrals are understood, here and below,
 to be principal-valued. Note that we use a nonstandard normalization
 in our definition of ${\cal H}$.}.\par In addition to satisfying the $1d$
 integro-differential equation, $\rho$ must be normalized to unity
 (by eq. (8)):
  $$\int d\lm\rho(\lm)=1\eqno(9d)$$
 The eq. (9b) reduces to the integral equation of ref.\biz\
  in the planar ($N\rightarrow\infty$) limit; the extra term can be
 physically understood as due to the variation of the entropy term
in the exponent of eq.(5a). Its effect is to smear the edges of the
 Dyson sea (or seas, for a multi-well potential $V$), so they become
 transition regions, rather than singularities as in the
 $D=0$ and $D=1$ collective field theories of Jevicki et. al. We will
 further discuss the transition regions in section 4
\REFS\bowick{ For a discussion of the transition regions using the
 standard Gel'fand-Dikii formalism, see: M.J. Bowick and E. Brezin
\journal Phys.Lett.&B268(91)21.}\refsend
 .\par
 For $N>>1$, it is easy to see from eq.(9b) that outside the sea, and
 beyond the transition regions, $\rho(\lm)$ is well-approximated by
$$\rho(\lm)\approx const\;\lm^{2N}e^{-N\ll V(\lm)+O(1/\lm)\rr}
\eqno (9e)$$ Thus, two conclusions can immediately be drawn.
 Firstly, a normalizable classical $\rho$ exists only if $V(\lm)$
 is bounded from below; and secondly, $\rho(\lm)$ vanishes nowhere,
 so that multiple eigenvalue bands (`seas') are related by a tunneling
 effect. This effect serves to determine the relative population of
multiple seas, and is responsible in general for string-nonperturbative
 effects.\par We note that, as our classical solution depends on $N$
 and hence includes some higher-genus effects, the terms `planar' and
 `classical' are not synonymous in our formalism. This is also the case
 for the Das-Jevicki field theory.
\par Once $\rho$ is known, the corresponding $I_s$ is determined
uniquely by eqs.(9); however, $I_s$ drops out of the formalism from
 here on.\par In the appendix we use eqs.(9b-d), and properties of
the Hilbert transform, to derive the following nonlinear Schr\"odinger
 equation:

    $$\bigl(-{1\over N^2}{\prt^2\over{\prt\lm^2}}+
    V_1(\lm)+\pi^2\rho^2\bigr)
        \bigl(\rho^{-1/2}\bigr)=0\eqno(10a)$$

where $V_1$, henceforth called the `Schr\"odinger potential',
 is the polynomial

$$ V_1(\lm)={1\over 4}(V^\prime)^2+{1\over{2N}}V^{\prime\prime}
       +P(\lm)\eqno(10b)$$

 $P$ is a polynomial whose coefficients are moments of the charge
distribution:

 $$P(\lm)\equiv -\int d\mu\rho(\mu){{V^\prime(\mu)-V^\prime(\lm)}\over
    {\mu-\lm}}\eqno(10c)$$
This Schr\"odinger equation is local, except for the coefficients of
$P(\lm)$, which are determined self-consistently. For example, in the
  case of quartic $V$, $$V(\lm)={1\over 2}
\lm^2+g\lm^4\eqno (10d)$$ the polynomial $P(\lm)$ becomes:
$$P(\lm)=-1-4g(m_2+\lm^2)\eqno (10e)$$ where the eigenvalue moments
 are defines as $$m_n=\int\lm^n\rho(\lm)d\lm\eqno (10f)$$
and we have used the symmetry $\rho(\lm)=\rho(-\lm)$, which follows
 from eq. (9b) and the symmetry of $V(\lm)$
\foot{For simplicity, we assume throughout that $V(\lm)=V(-\lm)$.
 Actually, this only implies a symmetric $\rho$ if $\rho(\lm)$ is
 unique, but we will ignore this complication here, since the entire
 analysis can be easily redone without the symmetry assumption, and
 besides the possible nonsymmetry will not affect perturbation
 theory.}
. The nonlinear Schr\"odinger equation has several points of interest,
 which we now discuss. Firstly, in it $1/N$ plays the role of Planck's
 constant, and the tunneling effects implied by eq.(9e), can now be
 seen to be similar to quantum mechanical tunneling. This is demonstrated
 in section 4, using the WKB approximation. In the exterior of the Dyson
 sea(s), $\rho$ is exponentially suppressed for large $N$, and (10a)
 approximates a linear Schr\"odinger equation. a strange feature of this
 correspondence is that the `wavefunction' here is $1/\sqrt{\rho}$,
the inverse of the intuitive $\sqrt{\rho}$. Another feature, probably
 closely related, is that the $V^{\pr\pr}$ term in our Schr\"odinger
potential $V_1$, has the opposite sign compared with the $D=1$ potential
 arising from the Marinari-Parisi reformulation\ \ll\marinar\rr\
\foot{The relation between the two facts is, that in order to get the
Marinari-Parisi $D=1$ Schr\"odinger equation, one chooses as wavefunction
 $\psi\{M\}=\exp(-{N\over 2}\tr V(M))$, with $M$ the original random
 matrix. Thus, if the Van der Monde
  were to be ignored, the one-particle
 wave-function would be the {\it square root} of $\rho$. If we choose
 the inverse wave function, the sign of $V^{\pr\pr}$ in the Schr\"odinger
 potential agrees with ours, rather than with that of Marinari
  and Parisi.}.
 \par The transition regions interpolate between the WKB solutions
outside and inside a given Dyson sea, similar to the role of the Airy
 function in ordinary (linear) WKB (see section 4). In the sea interior,
the $\rho^2$ term in (10a) is no longer negligible, and in fact
 there the leading WKB approximation is $$\rho(\lm)\approx{1\over\pi}
\sqrt{-V_1(\lm)}\quad(\; inside\; a\; sea\;)\eqno(10g)$$
 which is just the generalization of Wigner's semicircle law to
arbitrary potential $V$\ \ll\biz\rr\ .
\par In the planar limit and outside the Dyson sea, $V_1(\lm)$ has
a physical interpretation: $4V_1(\lm)$ is the square of the gradient of
 the effective potential, that is, the square of the total force exerted
 on a single eigenvalue, due to the potential $V(\lm)$ and the repulsion
 of the other $(N-1)$ eigenvalues.\par
As stated in the introduction, the Schr\"odinger equation is weaker
than the integro-differential equation. This is because (10a) is a
 second-order differential equation, so for given $V_1$ its solution
 $\rho(\lm)$ has two free continuous real parameters. By eqs.(10),
  $V_1$ has one additional unknown, $m_2$; but we must impose the
two self-consistency conditions, for $m_2$ and for $m_0=1$. This
leaves us with a {\it single undetermined real parameter}\foot
{Note that the usual WKB procedure for bound states, does not apply
for the Dyson sea. In the usual procedure, one imposes that the
wavefunction component which blows up exponentially at spatial infinity,
vanishes. But here the wavefunction is $\psi=\rho^{-1/2}$, so in fact
 $\rho$ normalizability {\it requires} $\psi$ to blow up exponentially,
 and no useful information results from this condition.}
. However,
 as is proven in the appendix, the  integro-differential equation
 allows no zero modes that preserve the normalization condition (9d).
 Hence, the free real parameter can only assume discrete values.
\chapter{Quantum Corrections.}
We next address the quantum corrections to $Z_{\scriptstyle classical}$
( eq. (5a)), needed in order to regain the full partition sum
 $Z$ in eq. (5). Let us separate the field $A$ into its classical and
quantum pieces,
 $$A(r)=-iA_s(r)+A_q(r)\eqno(11)$$
and also separate out the quadratic part of the action:
$$S\{A\}=S_{classical}+S_I\{A_q\}+{1\over 2}\int\int d^2rd^2r^\prime
A_q(r)K(r,r^\prime)A_q(r^\prime)\eqno(12)$$

Here $S_I$ is the interacting piece, consisting of the terms
of order three
and higher in $A_q$. $K$ is the inverse propagator of the quantum field
in the background of the classical solution:

$$K(r,r^\prime)=-(\prt_r)^2\delta(r-r^\prime)+4\pi N\delta(x)\delta
(x^\pr)\{\rho(\lm)\delta(\lm-\lm^\pr)-\rho(\lm)\rho(\lm^\pr)\}\eqno(13)$$
The only zero mode of $K$ is the constant function\foot
{This is equivalent to the fact, proven in the appendix, that
$\rho(\lm)$ has no normalization-preserving zero modes.}, so
 we fix that by defining our space of configurations $A_q$ to
satisfy $A_q(r)\rightarrow 0$ as $\vert r\vert\rightarrow\infty$. This
renders $K$ nonsingular\foot{This choice of boundary condition follows
from the fact that both $A(r)$ and $-iA_s(r)$ satisfy the boundary
 condition (7c).}.\par
Now, $S_I$ depends only on the value of $A_q$ on the $x=0$ axis, so we
define the one-dimensional field
$$q(\lm)\equiv\sqrt{4\pi}\bigl[A_q(\lm 0)-\int\rho(\mu)d\mu A_q(\mu 0)
\bigr]\eqno(14)$$
and denote
$$S_i\{q\}\equiv S_I\{A_q\}\eqno(15)$$
We then have, by eqs.(6),(6a),(8),(11),(12) and (14):
$$S_i\{q\}=-N\ln\bigl(\int\rho(\lm)d\lm
e^{iq(\lm)}\bigr)_{NQ}\eqno(16)$$
where $NQ$ denotes the non-quadratic part. The full partition
 function, including quantum corrections, is given by
 $$Z\{V\}=\sum Z_{\scriptstyle classical}\{V\}
 \bigl({{\det K}\over{\det(-\prt^2)}}\bigr)
 ^{-1/2}\langle
e^{-S_i\{q\}}\rangle\eqno(17)$$
where the expectation value is a Gaussian average, i.e. evaluated
 via Wick's theorem, the sum is over the discrete set of classical
solutions, the functional determinants are subject to the boundary
 condition $\lim_{\vert r\vert\rightarrow\infty}A_q(r)=0$, and for each
 classical solution, $Z_{classical}$ is given by eq.(5a).
\par The two-point function, to be used in the Wick expansion, is
$$\langle q(\lm)q(\lm^\pr)\rangle=4\pi\bigl(H(\lm,\lm^\pr)+
{1\over{4\pi N}}\bigr)\eqno (18)$$
where \hthing\  is the restriction to the eigenvalue axis of
 $H(r,r^\pr)$; the latter is defined uniquely by the following
properties ---
\smallskip  $$\prt^2_rH(r,r^\pr)=
4\pi N\rho(\lm)\delta(x)H(r,r^\pr)+\delta(r-r^\pr)\eqno (19a)$$
 $$H(r,r^\pr)=H(r^\pr,r)\eqno (19b)$$
 $$H(r,r^\pr)\approx H(\infty,r^\pr)+O({1\over{\vert r
\vert}}) \eqno (19c)$$
as $\vert r\vert\rightarrow\infty$, where $H(\infty,r^\pr)$ is a finite
 function of $r^\pr$.\par
  By integrating eq.(19a) over $r$ with measure $\int d^2r$ and using
 (19c), we find a fourth property:
$$\int\hthang\rho(\lm)d\lm=-{1\over{4\pi N}}\eqno (19d)$$
For actual calculations, only the one-dimensional restriction \hthing
\ is required. For $\lmp\approx\lm$, $H$ is
dominated by the logarithmic
singularity of the free $2d$ Green's function:
$$\hthang\apx{1\over{2\pi}}\ln\vert\lm-\lmp\vert+\; regular
\quad(\lmp\apx\lm)\eqno (19e)$$  \par
Equation (17) is our central tool for evaluating physical quantities
 using the conjugate-field formalism. We shall refer to the last,
Wick-expanded factor as the `Feynman diagrams', since they go beyond
the semiclassical determinant factor\foot{In the sense that they involve
 the interaction piece of the quantum action.}
. \par The only remaining divergences in eq.(17) are $UV$ ones, and
they appear in two places: the $\epsilon^{-N}$ factor in $Z_{classical}$
(see eq.(5a)), and the determinant factor
\foot{Superficially, it appears that the normal-ordering contributions
 to the Feynman-graph factor are also divergent, due to (19e); but
 these normal-ordering divergences manifestly cancel to all orders,
 as we shall show below.}~.
 But these two divergences cancel, as we now show. Using eq.(13)
and formally expanding in powers of the free massless propagator,
 we find:
$$\left.\eqalign{
\ln\bigl[{{\det K}\over{\det(-\prt^2)}}\bigr]^{-{1\over 2}}=&
-{1\over 2}\tr\ln\bigl[K/(-\prt^2)\bigr]\cr &= -2\pi N\int
d\lm\rho(\lm)\bigl({1\over{-\prt^2}}\bigr)_{\lm\lm}+\; UV\; finite
}\right.\eqno (20)$$
 This is ill defined. But if we use the $UV$ regularization eq.(2a),
 the $\delta(\lm-\lmp)$ term in $K(r,r^\pr)$ is smeared, and (20)
becomes
$$\ln\bigl[{{\det K}\over{\det(-\prt^2)}}\bigr]^{-{1\over 2}}=
N\ln\epsilon+\; UV\; finite\eqno (20a)$$
and hence the singular $\epsilon$-dependence drops out of eq.(17), as
claimed.
\par The remainder of this paper deals mostly with properties of the
 functions $\rho(\lm)$ and \hthing, especially in the double scaling
 limit, and with the summation of the Feynman graphs in eq.(17). The
 evaluation of $\det K$, which is done using heat-kernel methods, will
 be reported on elsewhere.\par To get an idea what the expressions for
 Feynman graphs look like, we record the lowest-order terms in the Wick
 expansion:
$$\left.\eqalign{
\ln\langle e^{-S_i\{q\}}&\rangle=\ln\langle\ll\int\rho(\lm)d\lm
e^{iq(\lm)}\rr^N\exp\bigl({N\over 2}\int\rho(\lm)d\lm q(\lm)^2\bigr)
\rangle\cr &=
{N\over 8}\{\int d\tau\ll\langle q(\tau)^2\rangle-\int d\tau^\pr\langle
q(\tau^\pr)^2\rangle\rr^2\cr &
-2\int\int d\tau d\tau^\pr\langle q(\tau)
q(\tau^\pr)\rangle^2+\dots\}
}\right.\eqno (21)$$
 In eq.(21), the leftmost equality is exact, and we have introduced a
 useful new variable $\tau(\lm)$:
$$\tau(\lm)\equiv\int_{-\infty}^\lm\rho(\mu)d\mu\eqno (21a)$$
which ranges from $0$ to $1$. We will sometimes denote the arguments
of a function of $\lm$, by $\tau$ instead
\foot{In deriving the expansion (21), we made use of the fact that
$\int d\tau q(\tau)=0$, by eqs.(14),(21a).}
. \par The terms displayed in the expansion on RHS of (21), are those
 involving only two propagators. Using eq.(18), we rewrite these
 Feynman-graph contributions to the free energy, thus:
$$\left.\eqalign{
\ln\langle e^{-S_i\{q\}}\rangle
=&2\pi^2N\{\int d\tau\ll H(\tau,\tau)-\int d\taup H(\taup,\taup)\rr^2
+{1\over{8\pi^2N^2}}\cr &
-2\int\int d\tau d\taup H(\tau,\taup)^2+O(H^3)\}
}\right.\eqno (21b)$$
The first term in the curly brackets is the normal-ordering contribution
 to this order. $H(\tau,\tau)$ is logarithmically divergent, but this
 divergence is $\tau$-independent, so $H(\tau,\tau)-\int d\taup
H(\taup,\taup)$ is $UV$ finite
\foot{This can be shown rigorously by making use again of the $UV$
regularization procedure, eq. (2a).}
{}.
 The other two-propagator graph in (21b) is manifestly finite. Note that
 the Feynman-graph contributions to the free energy are not local. Indeed,
 we started from a nonlocal action for the $A$ field, so this is to be
 expected -- the cluster expansion does not hold.\par
In order to evaluate Feynman graphs, we must solve for the propagator
$H$. Before doing so, however, let us show how to sum the normal-ordering
 contributions to all orders. This is easy to do: from the leftmost
equality in (21), we obtain
$$\langle e^{-S_i\{q\}}\rangle=\langle\{\int d\tau\exp(-w(\tau))
:e^{iq(\tau)}:\}^N\exp\ll{N\over 2}\int d\tau:q(\tau)^2:\rr\rangle
\eqno (22)$$
 where $w$ is a finite function, defined as follows:
$$w(\tau)\equiv w_1(\tau)-\int d\taup w_1(\taup)\eqno (22a)$$
$$w_1(\tau(\lm))\equiv\lim_{\lmp\rightarrow\lm}(2\pi\hthang-\ln\vert
\lm-\lmp\vert)\eqno (22b)$$
$w_1(\tau)$ is finite by virtue of eq.(19e).\par
We now turn to the task of computing \hthing. As was the case with
the $2d$ equation of motion (7), the two-dimensional differential
equation (19a) can be converted into a {\it one-dimensional}
integro-differential equation, by treating the RHS as a source term.
 This new equation is
$${\prt\over{\prt\lm}}\hthang={1\over{2\pi}}{1\over{\lm-\lmp}}+
2N{\cal H}_\lm\bigl(\rho(\lm)H(\lm,\lmp)\bigr)\eqno (23)$$
where the principal value is understood in the first term, and the
subscript to the Hilbert transform indicates that the transform
acts on $\lm$, with $\lmp$ held fixed. When the conditions (19b-c)
are imposed, eq.(23) has a unique solution.\par It is possible to derive
 from (23) a $1d$ {\it differential equation} for $H$, similarly to the
 procedure that lead from eqs.(9) to eqs.(10). Namely, eq.(23) is
 differentiated w.r.t. $\lm$, then used again, and the properties of
the Hilbert transform (listed in the appendix) are used. In addition,
eq.(9b) for $\rho$ is used. The result of these manipulations is
\foot{We use the $\tau$ variable, defined in (21a), with
$\taup=\tau(\lmp)$, $\tau=\tau(\lm)$.}:
$$\left.\eqalign{
\{{{\prt^2}\over{\prt\tau^2}}+4\pi^2N^2\}\hthang=&
-\pi N\delta(\tau-\taup)+{1\over{\rho^2(\lm)}}Q(\lm\vert\lmp)\cr&-
{1\over{2\pi\rho^2(\lm)}}{\prt\over{\prt\taup}}\bigl
({{\rho(\lmp)}\over{\lm-\lmp}}\bigr)
}\right.\eqno (24)$$
where principle value is again understood in the last term. Here
$Q(\lm\vert\lmp)$ is a polynomial in $\lm$, with coefficients that
 are one-sided moments\foot{That is, moments w.r.t. one of the two
 variables.}
 of $H$ with measure $\int d\tau$:
$$Q(\lm\vert\lmp)\equiv{N\over{2\pi}}{{V^\pr(\lm)-V^\pr(\lmp)}\over
{\lm-\lmp}}+2N^2\int\rho(\mu)d\mu{{V^\pr(\lm)-V^\pr(\mu)}\over
{\lm-\mu}}H(\mu,\lmp)\eqno (24a)$$
 and these moments are to be determined self-consistently, just as
 for the moments $m_n$ of $\rho$, which entered the
  Schr\"odinger eq.(10a) through the polynomial $P(\lm)$.\par
 Denote these one-sided moments as follows:
$$h_n(\mu)\equiv\int\lm^n\rho(\lm)d\lm H(\lm,\mu)\eqno (24b)$$
By (19d) we have, $$h_0(\mu)=-{1\over{4\pi N}}\eqno (24c)$$
Restricting again to the quartic potential, we obtain:
$$Q(\lm\vert\lmp)={1\over{2\pi}}4gN\{\ll\lmp^2+4\pi Nh_2(\lmp)\rr
+\lm\ll\lmp+4\pi Nh_1(\lmp)\rr\}\eqno (25)$$ \par
 The LHS and first term on RHS of (24) constitute the Green's equation
 for the correlator of a quantum-mechanical harmonic oscillator, but
 the other, nonlocal source terms on the RHS spoil this simple picture.
 As with the eigenvalue density, the $1d$ differential equation is
somewhat weaker than the $1d$ integro-differential equation: the latter,
 however, is equivalent to the {\it $2d$} Green's equation (19a), once
 the boundary condition (19c) is imposed.\par
The equation (24) is linear, and so can be readily solved in terms of
 $Q$ (which however is itself unknown). The general solution is:
$$\left.\eqalign{
\hthang=&A(\lmp)\cos 2\pi N\tau +B(\lmp)\sin 2\pi N\tau-{1\over 2}
\theta(\tau-\taup)\sin 2\pi N(\tau-\taup)\cr &+
{1\over{4\pi^2N}}\int_0^\lm{{d\mu}\over{\rho(\mu)}}\sin 2\pi N(\tau-
\tau(\mu))\ll2\pi Q(\mu\vert\lmp)+{\prt\over{\prt\taup}}\bigl(
{{\rho(\lmp)}\over{\lmp-\mu}}\bigr)\rr
}\right.\eqno (26)$$
 where $A$,$B$ are free functions and $\theta$ is the step function.
 The symmetry of \hthing\  determines $A$ and $B$ up to three real
 constants; these constants, as well as $Q(\lm\vert\lmp)$, can be
 determined by making use of eq.(23).\par
 The formalism for evaluating the classical solution(s), propagator
 and quantum corrections is rather complicated for
 finite $N$. Fortunately,
 however, massive simplifications occur in the ${1\over N}$ expansion,
 and the formalism simplifies even further when the couplings approach
 criticality and the d.s.l. (double scaling limit)
  is taken. We next turn to a discussion of this limit.
\chapter{The Double Scaling Limit.}
In this section, we will describe the procedure for performing the
 double scaling limit (\dasl ) in our conjugate-field formalism.
 We leave out many details, to be included in a forthcoming publication.
\par Let us specialize to the quartic potential, eq.(10d), and therefore
 to the $k=2$ critical model (pure gravity). The plan of the section is
 as follows. In part 4.a, the nature of the planar limit and the
  \dasl\  for
 the model is reviewed. Then the integro-differential and nonlinear
 Schr\"odinger equations for a classical solution, $\rho$, are used
 to find the string-perturbative expansions for the moments $m_n$ of
 $\rho$ (defined in (10f)). These expansions are unique, and it is seen
 they are unique for any critical potential, provided attention is
 restricted to classical solutions that are single-band in the planar
 limit. In addition, the WKB approximation for $\rho$ in the region
 exterior to the Dyson sea, is found.\par In part 4.b we discuss
 the perturbative corrections to $\rho$ inside the sea, and study the
 details of $\rho(\lm)$ in the transition regions at the edges of the
 sea. In part 4.c we describe how the same techniques, when applied
 to the propagator $H$, yield the perturbative expansion for \hthing
 \ in various regions of the $(\lm,\lmp)$ plane.
  The formulae of section
  3 for quantum corrections are seen to greatly simplify in the \dasl,
 allowing the perturbative series for physical quantities (specific
 heat, etc.) to be found. In particular, the normal-ordered
 Feynman diagram expansion terminates after a small number of terms.
\par Since the critical quartic potential is unbounded from below, all
 our results up to this point were obtained by continuing from the well
 defined $g>0$ regime, to $g\approx g_c<0$. This procedure is
 satisfactory only for string perturbation theory. In part 4.d, we
 look at the Schr\"odinger potential $V_1$ {\it directly} for
 $g\apx g_c$; $V_1$ is seen to acquire a small `second sea' in the
 transition region. This sea is finite in shape when $V_1$ and $\lm$
 are appropriately double-scaled.\par
  Our formalism is ill defined in this critical coupling regime,
  like pure gravity
 itself, since there is no normalizable $\rho(\lm)$. Nevertheless, we
 investigate the behavior of $\rho$ in the new transition region,
 and find that the second sea has a population suppressed by the
 expected $\exp(-const\;{1\over\kappa})$,
 where $\kappa$ is the string
 coupling. The constant in the exponential agrees with that obtained
in other approaches.
  Thus, $\rho$ can be thought of as an instanton. As such, it has
 two unusual aspects. Firstly, the tunneling factor
  occurs in the {\it field}
 configuration as well as in the classical action. Secondly,
  a single conjugate-field configuration seems to describe
  both the non-tunneling and tunneling
 eigenvalue configurations\foot{We thank S. Shenker for discussions
 concerning this latter point.}
 .  This configuration,
 however, should be viewed only as a warm-up exercise for non-perturbative
 calculations in our formalism. The meaningful calculations are to be
 done for a potential $V$ bounded from below, and would thus most likely
 not apply to any model with $k=2$ behavior\ \ll\deo\rr.\bigskip
\centerline{\bf 4.a. WKB and Perturbation Theory for $\rho$.}
{}From eqs.(10), we find the \shro\  potential for the case of quartic
 $V$\foot{Our notation, in particular our definition of $a(g)$, differs
 slightly from that of ref.\ \ll\biz\rr\ .}:
$$V_1(\lm)=4g^2(\lm^2-a^2)(\lm^2-b^2)^2-4g\delta m_2+{1\over{2N}}(1+
12g\lm^2)\eqno (27a)$$
where: $$a^2={1\over{6g}}(\sqrt{1+48g}-1)\eqno(27b)$$
$$b^2=-{1\over{4g}}-{1\over 2}a^2\eqno(27c)$$
and $\delta m_2$ is the deviation of the second moment $m_2$ from its
 planar-limit value:
$$\delta m_2\equiv
m_2+{1\over{36g}}-{{a^2}\over{144g}}(1+48g)\eqno(27d)$$
It will shortly be seen that $\delta m_2$ is $O({1\over N})$ for fixed
 $g$, in the planar limit.\par For $g>0$, $b^2$ is negative, so $V_1(\lm)$
 vanishes only at two points, which are $\lm=\pm a$ in the planar limit;
these points are the edges of the Dyson sea. We define $b$ such that
${\rm Im}\; b>0$. The $k=2$ critical point is at
$$g=g_c=-1/48\eqno(28a)$$
 and for $g\apx g_c$, $b^2$ is positive, and in
 fact $b^2\apx a^2\apx 8$
. As explained above, we will employ throughout most of section 4
 (except part 4.d) the well-defined procedure of calculating at $g>0$
 and then continuing to criticality. The double scaling limit for this
 theory, consists in simultaneously letting $N\rightarrow\infty$,
$g\rightarrow g_c$ while holding the string coupling fixed
\ \ll 1-8\rr\ :
 $$g_{string}=\kappa\equiv{1\over N}(g-g_c)^{-5/4}\eqno(28b)$$
\par Next, consider our nonlinear \shro\  equation, (10a), in a region
 of $\lm$ outside the sea; more precisely, $\vert\lm\vert>a$,
  and $\vert\lm\pm a\vert$ are held fixed as $N$ increases. The
 $\pi^2\rho^2$ (nonlinear) term is then exponentially suppressed, by
virtue of eq.(9e), and in consequence $\rho^{-1/2}$ approximately
satisfies a {\it linear} \shro\  equation:
$$\bigl[-{1\over{N^2}}{{\prt^2}\over{\prt\lm^2}}+V_1(\lm)\bigr]
(\rho^{-1/2})\apx 0\quad(\; outside\; sea\;)\eqno(29a)$$
This is easily solved via an asymptotic WKB expansion: ($V_1>0$ outside
 sea)
$$\left.\eqalign{
\rho(\lm)=&\vert\sqrt{V_1(\lm)}\vert\exp\bigl(-2N\int^\lm d\mu\sqrt
{V_1(\mu)}\bigr)\{const\; +O({1\over N})\cr &+
O(e^{\scriptstyle{
\scriptstyle -2N\int^\lm
d\mu\sqrt{V_1(\mu)}}})\} \quad(\; outside\; sea\;)
}\right.\eqno(29b)$$
where $V_1(\lm)$ is given by eq.(27a)\foot{It is easy to check that
(29b) agrees with eq.(9e).}
. The corrections on the RHS of eq.(29b) are of two kinds: the
 $O({1\over N})$ corrections constitute the usual, linear-WKB asymptotic
 expansion, whereas the exponentially-suppressed corrections are due to
 the nonlinearities of the exact (10a). Since our $V(\lm)$ is symmetric,
 so is $V_1$, and the branch we choose for $\sqrt{V_1(\mu)}$ in the
 exponent, is as follows:
$$ {\rm sgn}\sqrt{V_1(\lm)}\equiv{\rm sgn}\lm\eqno(29c)$$
This choice ensures that $\rho(\lm)$ is symmetric, and in
addition can be
 continued to an analytic function throughout the complex $\lm$ plane,
 except for a cut along the Dyson sea $(-a,a)$.\par
 Next, we take $\vert\lm\vert >>1$. The exponentially suppressed
corrections in (29b) can be neglected, and we obtain from eqs.(27a),(29b)
 an asymptotic expansion for the logarithmic derivative of $\rho$:
$$\left.\eqalign{
{{\rho^\pr(\lm)}\over{N\rho(\lm)}}\apx &{{2\delta m_2}\over{(\lm^2-b^2
)\sqrt{\lm^2-a^2}}}-{1\over{4gN}}{{1+12g\lm^2}\over
{(\lm^2-b^2)\sqrt{\lm^2-a^2}}}\cr &+{1\over N}({\lm\over{\lm^2-a^2}}+
{{2\lm}\over{\lm^2-b^2}})+O({1\over{N^2}})
}\right.\eqno(29d)$$
This expression can now be expanded as a Laurent series in ${1\over\lm}$,
and compared with the corresponding expansion resulting from eqs.(9)
 to yield the eigenvalue moments, $m_n$. The easiest way to compare the
 two series term by term, is to continue both of them to complex $\lm$,
 with $\sqrt{\lm^2-a^2}$ defined as described after (29c). We then
  multiply
both (9b) and (29d) by $\lm^n$, $n$ any integer, and integrate over $d\lm$
 along a closed contour with large $\vert\lm\vert$. For negative $n$,
 the coefficients agree identically. In addition, all odd moments vanish
 by symmetry\foot{This symmetry can be shown to hold to all orders
 in $1/N$ perturbation theory.}, whilst for even, positive $n$ values
 we find:
$$\left.\eqalign{
m_n=&{{4g}\over\pi}\int_{-a}^a\lm^n d\lm (\lm^2-b^2)\sqrt{a^2-\lm^2}+
{1\over{8\pi gN}}\int_{-a}^a\lm^n d\lm{{8gN\delta m_2-(1+12g\lm^2)}\over
{(\lm^2-b^2)\sqrt{a^2-\lm^2}}}\cr &+
{{a^n}\over{2N}}+d_0 b^n+O({1\over{N^2}})
\quad(\; n\; even\;)
}\right.\eqno(30a)$$
 with $$d_0={1\over{2N}}+{1\over{b\sqrt{b^2-a^2}}}({{\delta
 m_2}\over 2}
-{{1+12gb^2}\over{16gN}})\eqno(30b)$$
 For $n=0$ and $2$, this just reproduces eqs.(9d)
 and (27d) respectively,
 up to $O(1/N^2)$ terms, so we gain no new information. For higher $n$,
eqs.(30) give us all the moments in terms of $\delta m_2$, to order
 $1/N$. How is $\delta m_2$ to be determined, then? it is clear that
 $d_0$ must vanish to this order in $1/N$, since at $N>>1$ the support
of $\rho(\lm)$ is the Dyson sea, or very near it, and the $b^n$
 term on
the RHS of (30a) cannot occur for such a distribution. thus $d_0=
O({1\over{N^2}})$, which gives us the requisite missing information
\foot{The ${{a^n}\over{2N}}$ contribution in (30a), comes from the
edges of the sea; the remaining two terms come from its bulk.}
\foot{If $d_0\not=0$, ${\cal H}(\rho(\lm))$
must have poles at $\lm=\pm b$,
to first order in $1/N$. This is actually
 impossible for {\it any} $\rho(\lm)$
 for which the Hilbert transform is well-defined, since $b$ is
 imaginary.}
. ---
$$N\delta m_2={{1+12gb^2}\over{8g}}-b\sqrt{b^2-a^2}+O({1\over N})
\eqno(30c)$$
so $\delta m_2$ is indeed $O(1/N)$. Substituting this back into (30a-b)
yields all nonvanishing moments $m_n$, to order $1/N$. \par It is
straightforward to extend this technique to any desired order in
 $1/N$, and to take the \dasl\ limit (28b), as well. We thus see that the
 perturbative genus expansion for $\rho(\lm)$, or at least for the set
 of all its moments, is unique and can be easily determined, as claimed.
Furthermore, the technique extends to any potential $V$, as long as we
 have a planar limit of $\rho$ to expand about. When this limit is
 restricted to have a single band, it is unique, and so the perturbative
 expansion about it will also be unique.
\bigskip{\bf 4.b. Sea Interior and Transition Region.}
Consider the sea interior, namely the region $\vert\lm\vert<a$ with
 $a-\vert\lm\vert$ fixed (in either the planar- or the double-scaling
 limits). The nonlinear \shro\  equation (10a)
 can be solved via the WKB
 approximation in this region, as was done in the exterior region. In
 the interior, however, the linearized WKB is of no use. This is because
 the planar limit we wish to expand about is given by eq.(10g), and thus
 the nonlinearity is crucial here.\par
  The correct procedure is as follows.
 Defining new variables\foot{Recall that $V_1<0$ in the sea interior.},
$$t\equiv N\int_{-a}^\lm\sqrt{-V_1(\mu)}d\mu\eqno(31a)$$
$$\rho(\lm)\equiv{1\over\pi}\sqrt{-V_1(\lm)}f(t)^{-2}\eqno(31b)$$
 we find the differential equation,
$$f^{\pr\pr}+f-f^3=O({1\over{N^2}})\eqno(31c)$$
In the planar limit, the solutions of (31c) are elliptic functions.\par
These planar solutions have two free real parameters; this is just the
 ambiguity discussed at the end of section 2, and is resolved by the
 integro-differential equation and the consistency conditions. Let us
 see how this works. Since we expect $f(t)\rightarrow 1$ in the planar
 limit (by (10g) and (31b)), we need only consider $f\apx 1$; then (31c)
 informs us that
$$f(t)=1+\epsilon\cos 2\bar t+({3\over 4}-{1\over 4}\cos 4\bar t)
\epsilon^2+O(\epsilon^3)+O({1\over{N^2}})\eqno(32a)$$
where $\epsilon$ is a small unknown oscillation amplitude, and $\bar t=
t+\varphi$, with $\varphi$ the constant phase of the oscillation. By
employing the consistency conditions
 for $m_0=1$ and $m_2$, it can be shown
 that $$\epsilon=O({1\over N})\eqno(32b)$$ and this turns out to mean
 that the oscillations can be ignore in the \dasl\ ; thus we may use
$$f\apx 1\eqno(32c)$$ Next, we study the transition regions at the edges
 of the sea. By symmetry, it suffices to investigate the $\lm\apx a$
 transition region. Since $V_1(\lm)$ has a first-order zero at $\lm=a$
in the planar limit, we may approximate it by a linear function in the
 transition region, since the width of this region vanishes in the planar
 limit. We rescale $\lm$ and $\rho$ as follows:
$$\lm-a=N^{-2/3}\ll 8ag^2(b^2-a^2)^2\rr^{-1/3}y\eqno(33a)$$
$$\rho={1\over\pi}N^{-1/3}\ll 8ag^2(b^2-a^2)^2\rr^{1/3}\eta(y)^{-2}
\eqno(33b)$$
The eq.(10a) then becomes
$$(-{{\prt^2}\over{\prt y^2}}+y+\eta^{-4})\eta\apx 0\eqno(33c)$$
The boundary condition for this differential equation is furnished
 by the approximate solution inside the sea, eqs.(31b) and (32c), which
 become in terms of the rescaled variables,
$$\eta(y)\apx(-y)^{-1/4}\quad at\; y<0,\;\vert y\vert>>1\eqno(33d)$$
As $N\rightarrow\infty$, the rescaled equation (33c) becomes exact, and
(33d) becomes an exact boundary condition in the $y\rightarrow\infty$
 limit. On the exterior side of the transition region, the asymptotic
 behavior is
$$\eta(y)\apx const\; y^{-1/4}e^{(2/3)y^{3/2}} \quad at\; y>0,\;\vert
y\vert>>1\eqno(33e)$$
In agreement with the exterior WKB solution, (29b).\par
For fixed coupling $g$, the width of the transition region is $O(N^
{-2/3})$. When $g$ is, instead, continued to the critical point $g_c$
 in accordance with the \dasl, we find from eqs.(27)
$$b^2-a^2=O((g-g_c)^{1/2})\eqno(34a)$$
so by (33a) and (28b), the width of the transition region is of order
 $(g-g_c)^{1/2}\kappa^{2/3}$. Using eq.(33b), we also find that the
 normalized eigenvalue population in the transition region, is of order
 $$\int_{\scriptstyle transition}\rho d\lm=O({1\over N}),\eqno(34b)$$
 either for $g$ fixed, or in the \dasl. This means that of the original
 $N$ matrix eigenvalues, on the order of {\it one} eigenvalue are
 likely to inhabit the transition region.\bigskip
{\bf 4.c. Quantum Corrections in d.s.l.} In part 4.a, we used both the
 $1d$ differential equation and the integro-differential equation for
 $\rho(\lm)$, to find the perturbative expansion for the moments $m_n$
of $\rho$. A similar procedure can be employed for the propagator
 \hthing, by using the $1d$ Green's equation (24), and the corresponding
 integro-differential equation (23). In this case, one finds $1/N$
expansions for the one-sided moments $h_n(\lm)$, defined in (24b). The
 moments $h_1$ and $h_2$ can then be substituted in eq.(25). We find,
 to the leading approximation for large $N$,
 $$Q(\lm\vert\lmp)\apx{1\over{2\pi}}{{\prt}\over{\prt\taup}}\bigl(
\rho(\lmp){{\lm+\lmp}\over{b^2-\lmp^2}}\bigr)\eqno(35)$$
 This, in turn, can be used in eq.(26). The unknown functions $A$,$B$
 are determined as explained in section 3.\par
 The results of this analysis, are as follows\foot{The details will be
 presented in a separate publication, as will the computation of the
 determinant factor $\det K$.}. When $\lm$ and $\lmp$ are both interior
 to the sea, and $\lm\not=\lmp$, we have
$$\hthang=O({1\over N})\quad
\vert\lm\vert<a,\;\vert\lmp\vert<a\eqno(36a)$$
The function $w_1(\lm)$ (eqs.(22)), the regular piece of \hthing\ as
$\lmp\rightarrow\lm$, is $O(1)$, and in
 the sea interior it is approximated
 thus:
$$w_1(\lm)\apx\ln\rho(\lm)+\; const\quad(\vert\lm\vert<a)\eqno(36b)$$
In other regions of $\lm$ and $\lmp$, these functions have different
 behaviors. For instance, when $\lm$,$\lmp$ are both in the same
transition region,
$$\hthang=O(1)\quad(\;\lm\apx a,\;\lmp\apx a\;)\eqno(36c)$$
However, the contribution to Feynman diagrams from vertices in a
transition region, is still suppressed by a power of $1/N$ for each
such vertex, due to eq.(34b). Combining the behavior of $\rho$ and $H$
 in the various regions with eq.(22), we find that only the first few
 Feynman diagrams survive in the \dasl. We are referring to diagrams
 resulting from contractions among normal-ordered vertices; recall that
an infinite number of normal-ordering contractions have been summed to
 obtain eq.(22).
\bigskip{\bf 4.d. Instanton Configuration in Direct d.s.l.}
Rather than continuing $V(\lm)$ to criticality {\it after} solving for
$\rho$ and $H$, it is instructive to attempt taking the \dasl\ directly.
This will demonstrate what nonperturbative effects look like in the
 conjugate-field formalism, although a trustworthy nonperturbative
calculation requires a potential $V(\lm)$ bounded from below.\par
When we use the quartic potential (10d) with negative $g$, $b^2$ is
positive (part 4.a). As $g$ approaches $g_c$ from the origin along the
real axis ($g\apx g_c$, $g>g_c=-1/48$), $b$ approaches $a$ thus:
$$b>a\quad,\; b-a=O((g-g_c)^{1/2})\eqno(37)$$
 Thus by (27a), $V_1$ has two small `seas', concave regions just outside
the main Dyson sea, where it is negative. Unlike $V(\lm)$, $V_1$ {\it is}
bounded from below. Thus we can attempt to solve the nonlinear \shro
\  equation (10a) near criticality, ignoring the fact that the solution
will not solve the integro-differential equation\foot{This is similar to
the Marinari-Parisi approach to rendering the
 even-k models well-defined.}.
We shall continue to use eq.(30c) for $\delta m_2$, in eq.(27a). the
justification is that, assuming oscillations inside the sea are still
 suppressed (eq.(32c)), substitution of (31b) in (10f) for $n=0$ and $n=2$
 indeed yields (30c), at least to the approximation needed to take the
 \dasl\foot{This argument, as opposed to the one in 4.a, does not use
 information about large $\vert\lm\vert$. Such information cannot be
 trusted, as $V$ is unbounded from below there.}\foot{The transition
 region contributes $O(1/N)$ to both $m_0$ and $m_2$, so to eliminate
 these contributions we computed the subtracted $m_2-a^2m_0$.}
 .\par We concentrate on
 the behavior of $V_1(\lm)$,$\rho(\lm)$ in the new transition region;
  we invoke symmetry again and concentrate on the $\lm\apx a$ region.
To that end, we magnify the region via a new rescaling, different from
 (33):
$$\lm-2\sqrt{2}=4\sqrt{3}(g-g_c)^{1/2}x\eqno(38a)$$
$$\rho={1\over\pi}(g-g_c)^{3/4}\zeta(x)^{-2}\eqno(38b)$$
where $2\sqrt{2}$ is the value of $a$ at $g=g_c$. The \shro\ equation
 in this region assumes the form,
$$\bigl[-{1\over{48}}\kp^2{{\prt^2}\over{\prt x^2}}+\bigl({
{64\sqrt{8}}\over{\sqrt{3}}}(x+\sqrt{2})(x-{1\over{\sqrt{2}}})^2-\kp{
{2\sqrt{2}}\over{3^{1/4}}}\bigr)+
\zeta^{-4}\bigr]\zeta\apx 0\eqno(38c)$$
 This becomes exact in the \dasl. The old transition region occurs at
 $x+\sqrt{2}=O(\kp^{2/3})$, and is thus part of the new transition
 region.
  We are assuming that the string coupling
 $\kp$ is small, in order to isolate the leading nonperturbative
 effect.\par
To the right of the old transition region, $V_1$ is positive for
$$-\sqrt{2}<x<{1\over{\sqrt{2}}}$$
and $\rho$ is exponentially suppressed. The extra new minimum of $V_1$
is at $x=1/\sqrt{2}$, and the new sea surrounding it, where $V_1<0$,
has width $\Delta x=O(\kp^{1/2})$ and depth $O(\kp(g-g_c)^{3/2})$.\par
 The resulting \dasl\ solution for $\rho$ in the new transition region,
has the following properties. Between $x\apx -\sqrt{2}$ and $x\apx
{1\over{\sqrt{2}}}$, $V_1>0$ and $\rho$ tunnels in accordance with (29b).
$\rho$ is thus suppressed in the small new sea, with the WKB suppression
 factor (up to prefactors which can be found)
$$ \Lambda^2\equiv\exp\bigl(-{{4\sqrt{6}}\over{5\kp}}(48)^{5/4}\bigr)
\eqno(39a)$$
which is precisely the tunneling factor appearing in the nonperturbative
 ambiguity for pure gravity, using the various previous approaches
\ \ll\npt\rr
 \foot{In this connection, note that our normalization
 for $\kappa$ differs from the one usually employed in the literature.}
.\par In the new sea, we find that $\zeta(x)$ has the following form:
$$\zeta(x)\apx ({3\over\kp})^{1/4}\sqrt{8}e^{-2\sqrt{2}z^2}\bigl(
d_1\Lambda^{-1}-d_2\kp^{1/2}\Lambda\int_0^ze^{4\sqrt{2}\bar z^2}d\bar z
\bigr)\eqno(39b)$$
where $z$ is yet a {\it third} rescaled eigenvalue variable, appropriate
 for the extra sea:
$$x-{1\over{\sqrt{2}}}\equiv\kp^{1/2}(48)^{-3/8}z\eqno(39c)$$
When the \dasl\ limit is taken, $\kp$ small and $z$ held fixed, the
corrections to (39b) are higher powers of $\kp$. In eq.(39b), $d_1$,$d_2$
 are two pure numbers, obtained by matching (39b) at large and negative
 $z$ with the WKB approximation, (29b), which holds between the two seas.
 \par The configuration given by eqs.(38-39) can be interpretted as an
 instanton. For well-defined models, there exist similar
 instantons which are true classical solutions of the conjugate-field
 theory. \par In the configuration discussed above, the center $z=0$
 of the new sea is a local minimum of $\rho$. $\rho$ then increases with
 $z$ for $z>0$. When $z$ is sufficiently large, one exits the new
  sea and enters another $V_1>0$
 region, where the exponentially suppressed terms in eq.(29b)
 dominate for a while; this allows $\rho$ to continue to increase.
  Eventually the dominant term will again dominate, but if $\rho^2$
 has by then increased to become comparable in magnitude to $V_1$,
 a nonperturbative solution of (10a) is needed. We need not worry about
 this exterior region, however, since in this model $\rho$ is not
a trustworthy configuration there.\bigskip
\chapter{Conclusions.}
We have presented a new field-theory formulation of $D=0$ matrix
 models. The field is conjugate to the Jevicki-Sakita collective field,
 i.e. conjugate to the density of matrix eigenvalues. The theory is
 two dimensional, with an eigenvalue
  coordinate and an auxiliary coordinate
 that can be eliminated from the formalism. The action is nonlocal, but
 the equation of motion is local, except for a self-consistency
 condition. There is a unique or discretely labeled classical solution
 for any well-defined potential. The equation for the classical eigenvalue
 distribution is a modified version of the planar integral equation
 of Bessis, Itzykson and Zuber, with an entropy term that smoothes the
 edges of the Dyson sea and introduces higher-genus
 effects already at the
classical level. Single-band classical solutions
 are perturbatively unique. The classical distribution also satisfies
 a nonlinear \shro\  equation,
 with a potential similar to, but different
 from the one appearing in the Marinari-Parisi $D=1$ reformulation.
 \par The classical solutions, and the
 quantum corrections about them, are systematically calculable,
 and all divergences (UV- and IR-) cancel manifestly to all orders.
  The
 normal-ordering graphs can be summed exactly to all orders. In the double
 scaling limit, the formalism simplified drastically. \par
 The conjugate-field formalism can be used to systematically compute
 string-nonperturbative effects. We demonstrate this for the ill-defined,
 but simple, case of $k=2$ realized with a
  quartic potential. In this case,
 the classical distribution contains two small seas on either side of the
 main Dyson sea. The population of the new seas
  is exponentially suppressed
 by the same tunneling factor as appears in other approaches.
A more complete presentation of the conjugate-field formalism will
 be presented in a forthcoming publication.\bigskip
\appendix
In this appendix, we derive the \shro\  equation (10) from the
integro-differential equation, eq.(9b), and prove that a classical
 solution does not have any normalization preserving zero modes.\par
First, we list a few useful properties of the Hilbert transform (which
in our normalization is given by eq.(9c)).
 For any function $f(\lm)$,
$${\cal H}(f^\pr(\lm))={d\over{d\lm}}{\cal H}(f(\lm))\eqno(A.1)$$
 Also, for any two functions $u$ and $v$, one easily finds\foot{
 By using
 the fact that the real and imaginary parts of an analytic function,
 having all its singularities in the lower half
  of the complex plane, are
 related by the Hilbert transform.}:
$${\cal H}(u{\cal H}(v)+v{\cal H}(u))={\cal H}(u){\cal H}(v)-\pi^2 uv
\eqno(A.2)$$
Actually, for the present derivation, we only need the degenerate form
 of this identity when $v=u$:
$$2{\cal H}(u{\cal H}(u))={\cal H}(u)^2-\pi^2 u^2\eqno(A.3)$$
But the more general (A.2) is needed to derive the $1d$ differential
 equation for the propagator \hthing\ , eq.(24), from the corresponding
 integro-differential equation (23). We will not go through this latter
 derivation, but the procedure parallels that for $\rho$.\par
Differentiating eq.(9b) w.r.t. $\lm$ and using (A.1), we obtain
$$V^{\pr\pr}+{d\over{d\lm}}({1\over N}{{\rho^\pr}\over\rho})=
2{\cal H}(\rho^\pr)\eqno(A.4)$$
 Now we use (9b) again to rewrite $\rho^\pr$ on the RHS of (A.4), and
 also use eq.(A.3). Upon rearranging terms and using the definition
 (9c), the nonlinear \shro\  equation, given by eqs.(10), is obtained.
\par Next, let $\delta\rho$ be an infinitesimal zero-mode of $\rho$
 which preserves the normalization condition (9d). Thus
 $$\int d\lm\delta\rho(\lm)=0\eqno(A.5)$$
and by varying (9b) we find a linear integro-differential equation for
$\rho$:
$${1\over N}\bigl({{\delta\rho}\over\rho}\bigr)^\pr=2\ch(\delta\rho)
\eqno(A.6)$$
 Let us define the function  $\eta(r)$ in the two-dimensional space,
 as follows: on the eigenvalue axis, it is defined as
$$\eta(\lm)\equiv{{\delta\rho}\over\rho}\eqno(A.7)$$
 and off that axis, it is defined by continuation of (A.6):
$$\eta(r)=const\; +2N\int\eta(\mu)\rho(\mu)\ln\vert r-r^\pr\vert d\mu
\eqno(A.8)$$
where $r^\pr=(\mu,0)$ runs over the eigenvalue axis in the integral.
 The additive constant in eq.(A.8) is determined by the condition
 (A.5), which we rewrite as
$$\int\eta(\lm)\rho(\lm)d\lm=0\eqno(A.9)$$
{}From (A.8) and (A.9) we find that
$$\eta(r)\rightarrow const\; +O({1\over{\vert r\vert}})
\quad as\; \vert r\vert\rightarrow\infty\eqno(A.10)$$
But by eq.(A.8), $\eta$ satisfies the $2d$ linear differential equation
$$\prt^2\eta(r)=4\pi N\delta(x)\rho(\lm)\eta(r)\eqno(A.11)$$
 where $r=(\lm,x)$. This differential equation has a unique solution
 subject to the boundary condition (A.10). We have thus proven that
 a classical solution has no normalization-preserving zero modes.

\ACK{
I would like to thank Adrian Cooper, Lenny Susskind and Larus
 Thorlacius for useful discussions and comments, and Stephen
 Shenker for discussions concerning instantons and eigenvalue
 tunneling.}
\refout
\bye